\documentclass[journal,twocolumn,final]{IEEEtran}
\usepackage[cmex10]{amsmath}
\usepackage{amssymb}
\usepackage{array}
\usepackage{mdwmath}
\usepackage{mdwtab}
\usepackage{eqparbox}
\usepackage{setspace,citesort}
\usepackage{pgf,pgfarrows,pgfnodes,pgfautomata,pgfheaps,pgfshade}
\usepackage{tikz,pgflibraryshapes,pgflibrarysnakes}
\usepackage{graphicx}
\usetikzlibrary{positioning,patterns,fadings}
\usepackage{electComp}
\usetikzlibrary{decorations,decorations.pathmorphing,decorations.pathreplacing}
\usepackage{array,txfonts}
\usepackage{tikz}
\usepackage{pgffor}
\usepackage{gensymb}
\usepgflibrary{plotmarks}
\usetikzlibrary{decorations.pathreplacing} 
\usepackage{subfig}
\let\MYoriglatexcaption\caption
\renewcommand{\caption}[2][\relax]{\MYoriglatexcaption[#2]{#2}}
\let\MYorigsubfloat\subfloat
\renewcommand{\subfloat}[2][\relax]{\MYorigsubfloat[]{#2}}

\graphicspath{{graphics//}}

\newcommand{\eps}{\varepsilon}
\newcommand{\bml}{\begin{subequations}}
\newcommand{\eml}{\end{subequations}}
\newcommand{\br}{{\bf r}}

\begin{document}
\title{Stability Properties of the Time Domain Electric Field Integral Equation Using a Separable Approximation for the Convolution with the Retarded Potential}

\author{{A.~J.~Pray,~\IEEEmembership{Student Member,~IEEE,} N.~V.~Nair,~\IEEEmembership{Member,~IEEE,}~and~B.~Shanker,~\IEEEmembership{Fellow,~IEEE,}}
\thanks{The authors are with the Department of Electrical and Computer Engineering, Michigan State University, East Lansing, MI, USA, 48824 e-mail: prayandr@msu.edu .}
}

\maketitle

\begin{abstract}
	The state of art of time domain integral equation (TDIE) solvers has grown by leaps and bounds over the past decade. During this time, advances have been made in (i) the development of accelerators that can be retrofitted with these solvers and (ii) understanding the stability properties of the electric field integral equation. As is well known, time domain electric field integral equation solvers have been notoriously difficult to stabilize. Research into methods for understanding and prescribing remedies have been on the uptick.  The most recent of these efforts are (i) Lubich quadrature and (ii) exact integration. In this paper, we re-examine the solution to this equation using (i) the undifferentiated form of the TD-EFIE and (ii) a separable approximation to the spatio-temporal convolution. The proposed scheme can be constructed such that the spatial integrand over the source and observer domains is smooth and integrable. As several numerical results will demonstrate, the proposed scheme yields stable results  for long simulation times and a variety of targets, both of which have proven extremely challenging in the past. 

\end{abstract}

\section{Introduction}
Since the development of TDIE based methods in the 1960's \cite{Friedman1962}, computational complexity and late time 
instability have been the two stumbling blocks that have prevented their widespread adoption as electromagnetic analysis
tools. Over the past decade, the former has been largely addressed and fast evaluators with provable error
estimates that can be retrofitted with TDIE solvers exist \cite{Yilmaz2001,Shanker1999a}. In fact, TDIE solvers augmented with these fast
solvers have seen widespread application to a range of electromagnetics and acoustics applications \cite{Shanker1998a,Ergin2000a}. While 
the issue of computational complexity is a solved problem, that of instability still lingers.  Over the years there have been several attempts at stabilizing
these equations; these include filtering techniques \cite{Sadigh1993}, implicit time stepping \cite{Sarkar2000}, smooth basis functions
\cite{Hu2001}, space-time Galerkin methods \cite{Abboud2001}, and Lubich quadrature \cite{Wang2008}. Methods that are based on transform techniques have been developed as well \cite{Sarkar2000}, and they avoid problems with instability by avoiding time marching
altogether. As they avoid marching on in time (MOT), they will not be the subject of the ensuing discussion. 

Despite the plethora of interest, instability of TDIE solvers is far from being a solved problem. Of all the methods
listed above, only the space-time Galerkin methods are provably stable \cite{Ha-Duong2003}. Here, stability is proven by passage through the Fourier-Laplace domain. Indeed, most of the others, with the possible exception of Lubich quadrature, simply delay the onset of instability. 
While the stability properties of space-time Galerkin based schemes have been proven, such methods are notoriously
difficult to implement. These methods
rely on two aspects to obtain stability; (i) construction of the TDIE based on a variational formulation and (ii) exact
evaluation of all integrals involved. The latter involves the exact evaluation of five-dimensional integrals
\cite{Abboud2001}. However, as most of literature on this topic is published only as Ph.D theses, only
sparse details have been available to the rest of the scientific community. Recent papers have sought to address some
of these concerns \cite{Ha-Duong2003}, but are largely restricted to problems in acoustics. In the electromagnetics community,
attempts to develop such a scheme started with the development of space-time collocation together with exact evaluation
of fields for both scattering from perfectly conducting and composite \cite{Shanker2009} objects. It was shown that
this method was indeed stable for a whole class of targets that had, in our experience, been impossible to stabilize
with any of the other existing techniques. More recently, a space-time Galerkin formulation for electromagnetics was introduced
\cite{Shi2011}, wherein stability was shown for extremely long solution times, small time steps and higher
order temporal interpolants. 

The key in implementing this method is determining the topology of the domain of integration. This is best illustrated
by considering the scheme within a collocation framework as done in \cite{Shanker2009}. Here, to test the field 
received by a triangular patch due to a point source whose time signature is piecewise continuous, one has to find 
the corresponding domains in the triangle where the integrands are piecewise continuous. This is tantamount to finding
intersections of the triangle with concentric time spheres of radii $i c \Delta_t$ that are centered at the source
point, where $c$ is the speed of light, $\Delta_t$ is time step size, and $i$ is an integer. It is readily apparent 
that while this method is very effective, it is impossible to use
in a higher order framework and extremely difficult to integrate with fast methods. Another approach that has
seen recent attention is the application of Lubich quadrature \cite{Wang2008} to TDIEs. This method relies on a
series of transforms, to the Laplace domain, the $z$-domain, and back into the time domain.
As it is based on a series of domain transformations, it is possible to analyze
propagation through and scattering from dispersive and lossy materials. On the flip side, as it is based on
entire-domain transforms, the method converts what should be an ${\cal O}(N_t N_s^2)$ solver given the compact
nature of the retarded potential to a system whose complexity scales as ${\cal O}(N_t^2 N_s^2)$. Here, $N_t$ and 
$N_s$ are the number of temporal and spatial degrees of freedom, respectively. To overcome this bottleneck, one
would then have to take recourse to an FFT based method \cite{Hackbusch}.

The goal of this paper is to present a method that yields stable results for the time domain electric
field integral equation (TDEFIE) while obviating some of the aforementioned drawbacks. It is based on the undifferentiated
form of the TDEFIE (and to a large extent inspired by \cite{Shanker2009,Shi2011}), and relies on deriving an alternate representation
of convolution between the retarded potential and the space-time basis function. Key attributes of this method are
the following: (i) it does not transform the problem to the Laplace domain, (ii) it has finite support in time, and
(iii) it is entirely numerical, and as a result, it can be extended to higher order discretizations (in space and time) and integrated
with existing fast methods. The principal contributions of this paper are threefold: 
\begin{enumerate} 
\item We will present a methodology based on a separable approximation to the space-time convolution of a source with the retarded potential. We will elucidate its implementation within an MOT framework. 
\item We will frame the MOT system as an eigenvalue problem (akin to that done in the \cite{Walker2002} but for the system of equations used here). This will be used to provide insight into the stability of the system for a given discretization and time step size. 
\item Finally, to demonstrate the effectiveness of the method, we will present scattering results from a number of 
	benchmark targets, with RCS comparisons to a validated frequency domain solver or analytical results where possible.  A number of the presented targets
	have defied all stabilization schemes apart from \cite{Wang2008,Shanker2009,Shi2011}.
\end{enumerate}
In this paper, we shall not present results of extension of this method to either higher order surfaces or integration
with fast solvers. Likewise, we will rely on demonstration of stability via a set of challenging targets as opposed to
mathematical proofs for late time stability. 

The rest of this paper is organized as follows: Section \ref{Section:formulation} details the approach used in this
paper, contrasts it with methods used elsewhere. Implementation details of this approach are presented in Section
\ref{Subsection:implementation}. Details of the eigenvalue analysis implementation is presented in Section \ref{Section:eigen}.  Section \ref{Section:results} presents a number of results that demonstrate late time 
stability.  Lastly, Section \ref{Section:summary} summarizes the contribution of this paper and outlines directions
of future research.

\section{Formulation\label{Section:formulation}}
Consider a perfectly electric conducting body occupying a domain $D_{-} \subset \mathbb{R}^{3}$ in free space ($\mu_0,\eps_0$). Let $\Omega$ denote the bounding surface of $D_{-}$ and let the outward pointing unit vector normal at any point ${\bf r}$ on $\Omega$ be denoted by $\hat{n}({\bf r})$. Assume that a field $\{ {\bf E}^i({\bf r},t) \mbox{, } {\bf H}^i({\bf r},t) \}$ incident on this body is bandlimited to frequency $f_{max}$ and is zero for $t<0$.  This field induces currents, ${\bf J}({\bf r},t)$, on the surface $\Omega$, that radiate scattered fields  $\{{\bf E}^s({\bf r},t) \mbox{, } {\bf H}^s({\bf r},t) \}$. Thus, the total electric and magnetic fields in all of $D_{+} = \mathbb{R}^3/D_{-}$, denoted by ${\bf E}({\bf r},t)$ and ${\bf H}({\bf r},t)$, can be decomposed into the incident and scattered fields as 
\bml
\label{eq:flds}
\begin{equation}
\begin{split}
{\bf E} ({\bf r},t)  & = {\bf E}^i ({\bf r},t) + {\bf E}^s({\bf r},t)~, \\
{\bf H}({\bf r},t) & = {\bf H}^i ({\bf r},t) + {\bf H}^s ({\bf r},t)~.
\end{split}
\end{equation}
The currents induced can be solved for using either the electric, magnetic or combined field integral equation.  The TDEFIE has historically been the most challenging to stabilize.  Therefore, the rest of this paper will focus on discretizing this equation.
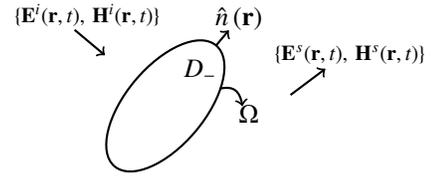
\begin{figure}[!h]
     \centering
\begin{tikzpicture}[scale=0.35,thick]
    \begin{scope}[scale=0.75,thick]
    \pgftransformrotate{-40}
        \draw(0,0) ellipse (2cm and 4cm);
    \draw[->](1.6,2.5) .. controls (2.0,3.0) and (2.4,3.0) .. (3.0,2.5);
    \draw[->](0.0,4.0)--(0.0,5.0);
    \draw (0.0,5.8) node {$\hat{n}\left({\bf r}\right)$};
    \draw(3.5,2.4) node {$\Omega$};
    \draw(0.2,2.5) node {$D_-$};
    \draw[->](-6,0)--(-4,0);
    \draw(-6,1) node {\footnotesize $\{{\bf E}^i({\bf r},t)$, ${\bf H}^i({\bf r},t)\}$};
    \draw[->](4.5,4.5)--(5.0,6.65);
    \draw(5.5,8.0) node {\footnotesize$\{{\bf E}^s({\bf r},t)$, ${\bf H}^s({\bf r},t)\}$};
    \end{scope}
\end{tikzpicture}
\caption{General description of a scattering problem\label{fig:generic}}
\end{figure}
The TDEFIE may be written as
\begin{equation}
\label{eq:IntEq} 
\begin{split}
	\hat{n} \times\hat{n} \times &{\bf E}^i({\bf r},t) = - \hat{n} \times \hat{n} \times{\bf E}^s \circ \left \{ {\bf J}({\bf r}, t) \right \} ~ \forall {\bf r}\in \Omega~,\\
{\bf E}^s\circ \left \{ {\bf J}({\bf r}, t) \right \}  &= - \partial_t {\bf A} \circ \left \{ {\bf J}({\bf r}, t) \right \} - \nabla \Phi \circ \left \{ {\bf J}({\bf r}, t) \right \}~,\\
{\bf A} \circ \left \{ {\bf J}({\bf r}, t) \right \} & = \frac{\mu_0}{4 \pi}  \int_\Omega d{\bf r}' \frac{{\bf J}({\bf r},\tau)}{R} ~,\\
\Phi \circ \left \{ {\bf J}({\bf r}, t) \right \} & =  \frac{1}{4 \pi \eps_0}\int_\Omega d{{\bf r}'} \int_{-\infty}^\tau dt' \frac{\nabla' \cdot {\bf J}({\bf r}', t')}{R} ~,
\end{split}
\end{equation}
\eml
where $R = |{\bf R}| = |{\bf r} - {\bf r}'|$, $\tau = t - R/c$ and $\nabla'$ denotes a divergence with respect to ${\bf r}'$. The solution to this integral equation for the unknown currents ${\bf J} ({\bf r},t)$ is typically effected by representing the current in terms of space-time basis functions as 
\begin{equation} 
\label{eq:SpaceTimeBasis}
{\bf J}({\bf r},t) = \sum_{n=1}^{N_s} {\bf S}_n ({\bf r}) \sum_{i=1}^{N_t} J_{n,i}T_i(t)~,
\end{equation}
where the temporal coefficients $J_{n,i}$ are to be determined for $N_t$ temporal and $N_s$ spatial basis functions. In the above expressions, the spatial basis functions ${\bf S}_n ({\bf r})$ are chosen to be the Rao-Wilton-Glisson basis functions \cite{Rao1982} defined for each edge on the tesselation that describes $\Omega$ such that 
\begin{equation}
{\bf S}_n({\bf r}) = \left \{ \begin{array}{cc}
\dfrac{l_n}{2 A_n^+} \left ( {\bf r} - {\bf r}_n^+ \right ) & {\bf r} \in P_n^+ \\
- \dfrac{l_n}{2 A_n^-} \left ( {\bf r} - {\bf r}_n^- \right ) & {\bf r} \in P_n^- 
\end{array}
\right . ~,
\end{equation}
where $P_n^\pm$ are the two triangles (of areas $A_n^\pm$, respectively) that are associated with each edge $n$, $l_n$ is the length of the edge, and ${\bf r}_n^\pm$ are free vertices associated with $P_n^\pm$. The spatial basis function vanishes outside the domain $\Omega_n = P_n^+ \cup P_n^-$. The temporal basis functions used, $T_i (t) = T(t- i \Delta_t)$, are typically $p$th order shifted Lagrange polynomials given by \cite{Shanker2009}
\begin{equation}
	\begin{split}
		T(t) =& f_k (t) g_{p-k}(t) {\cal P}_{\alpha, \beta} (t)~,\\
		& \alpha = (k-1)\Delta_t; \beta =  k\Delta_t \text{ for } k = 0, \cdots, p~,
	\end{split}
\end{equation}
where ${\cal P}_{\alpha,\beta}(t)$ is a rectangular pulse function in the domain $[ \alpha, \beta ]$, 
\begin{equation}
f_k (t) = \left \{
\begin{array}{cc}
1 & k = 0 \\
 \displaystyle \prod_{j = 1}^k \dfrac{t-j \Delta_t}{(-1)j \Delta_t} & k \ne 0
\end{array}
\right .~,
\end{equation}
and 
\begin{equation}
	g _{p-k} = \displaystyle \prod_{j = 1}^{p-k} \frac{t + j \Delta_t}{j \Delta_t }~,
\end{equation}
where $\Delta_t = \chi/(20 f_{max})$ is the time step size and $\chi$ is an oversampling factor.  Traditional MOT schemes are derived by substituting \eqref{eq:SpaceTimeBasis} in \eqref{eq:IntEq}, using Galerkin testing in space and point testing in time. The resulting equations may be succinctly written as 
\bml
\begin{equation}
\label{eq:MOT}
{\cal Z}_0 {\cal I}_j = {\cal F}_j - \sum_{i=1}^{j-1} {\cal Z}_i {\cal I}_{j- i} - \sum_{i=1}^{j-1}\tilde{\cal Z}_i {\cal C}_{j-i}~,
\end{equation} 
where 
\begin{equation}
{\cal I}_j = \left [J_{1,j}, J_{2,j}, \cdots, J_{N_s,j} \right ]^T,
\end{equation}
\begin{equation}
\label{eq:oldchargecurrel}
  {\mathcal C}_j = ~{\mathcal C}_{j-1} + \sum_{i = j-p-1}^{j-1} {\mathcal I}_{i} \int_{(j-1)\Delta t}^{j\Delta t} dt' T_{i}(t')~,
\end{equation}
\begin{equation}
{\cal F}_{n, j} = \left.\left < {\bf S}_n({\bf r}) , \hat{n} \times\hat{n} \times {\bf E}^i ({\bf r},t)\right >\right |_{ t= j \Delta_t}~,
\end{equation}
and
\begin{equation}
\label{eq:InnerProd}
\begin{split}
	{\cal Z}_{nm,i} =& \left.\left < {\bf S}_n({\bf r}) , \hat{n} \times\hat{n} \times {\bf E}_1^s \circ \left \{{\bf S}_m({\bf r})T_{j-i}(t) \right\} \right >\right |_{ t= j \Delta_t}~, \\ 
\tilde{\cal Z}_{nm,i} =& \left.\left < {\bf S}_n({\bf r}), \hat{n} \times\hat{n} \times {\bf E}_2^s \circ \left \{ {\bf S}_m({\bf r})T_{j-i}(t) \right \} \right >\right |_{ t= j \Delta_t}~, \\ 
{\bf E}^s =&~{\bf E}_1^s + {\bf E}_2^s~,
\end{split}
\end{equation}
where ${\bf E}_{1,2}^s$ are defined as
\begin{equation}
	\label{eq:phisplit}
	\begin{split}
		{\bf E}_1^s  \circ \left \{ {\bf S}_mT_{i} \right\}= &  - \partial_t {\bf A}_m \circ \left \{ {\bf S}_mT_{i} \right \} - \nabla \Phi_1 \circ \left \{ {\bf S}_mT_{i}  \right \}~,\\
		{\bf E}_2^s  \circ \left \{ {\bf S}_mT_{i} \right\} = & - \nabla \Phi_2 \circ \left \{ {\bf S}_mT_{i} \right \}~,\\
		\Phi_1 \circ \left \{ {\bf S}_mT_{i} \right \} = & \frac{1}{4 \pi \eps_0} \int_\Omega d{{\bf r}'}  \frac{\nabla' \cdot {\bf S}_m({\bf r}')}{R}
		\int_{k\Delta_t}^\tau dt' T_{i}(t')~,\\
		\Phi_2 \circ \left \{{\bf S}_m\right \} = & \frac{1}{4 \pi \eps_0} \int_\Omega d{{\bf r}'} \frac{\nabla' \cdot {\bf S}_m({\bf r}')}{R}~,
	\end{split}
\end{equation}
\eml
where $k = \lfloor \tau/\Delta_t\rfloor$.  In the above equations, $\left < \cdot \right >$ denotes a standard inner product. As is evident from \eqref{eq:MOT}, the solution proceeds sequentially over each time step. The crux of the solution of late time instability has been the accurate evaluation of \eqref{eq:InnerProd} (together with the use of the undifferentiated TDEFIE). It is evident from the nature of the temporal basis functions that the integrand in \eqref{eq:InnerProd} is piecewise-continuous, and the use of Gauss quadrature to evaluate these integrals before first identifying domains of continuity will be inaccurate.
 It should be noted that the situation is far worse when one uses the derivative form of the TDEFIE as the vector potential term
includes the second derivative of the temporal basis.  In what follows, we will seek to develop an alternate method for evaluating this integral. 

\subsection{Approximation of the temporal convolution}

\begin{figure}[h!]
	\begin{center}
       \begin{tikzpicture}[decoration=brace]
	       \draw (-1.75,-2) -- (-1,1) node[above] {\small \bf Source };
	       \draw (-1,1) -- (0.25,-0.5);
	       \draw (0.25,-0.5) -- (-1.75,-2);
	       \draw (2.7,1.45) -- (5,-.05) node[right,midway] {\small \bf  $~ ~ ~$  Observation};
	       \draw (5,-.05) -- (3,-.55);
	       \draw (3,-.55) -- (2.7,1.45);
	       \draw[->] (-.7,-.5) -- (3.84254,-.0457413) ;
	       \draw[<->] (1.67799cm,-0.49cm) -- (2.902cm,-.4cm) node[below,midway] {\small $c\Delta t$};
	       \draw[decorate] (-.8,-.4) -- (2.8575,-.05) node[above,midway] {\small $\zeta$} ;
	       \draw[decorate]  (2.963,-0.0299) -- (3.843,.0543) node[above,midway] {\small $R-\zeta$} ;
		\draw (.36527cm,-.885255cm) arc (-18.295:26.705:1.2273cm);
		\draw (1.53054cm,-1.27051cm) arc (-18.295:26.705:2.4546cm);
		\draw (2.75454cm,-1.65577cm) arc (-18.295:26.705:3.68191cm);
		\draw  (-.8cm,-.7cm) node[left] {\small \bf ${\bf r}'$}; 
		\draw  (3.94cm,-.02cm) node[right] {\small \bf ${\bf r}$}; 
	       \pgfplothandlermark{\pgfuseplotmark{x}}
		\pgfplotstreamstart
		\pgfplotstreampoint{\pgfpoint{-.8cm}{-0.5cm}} node[below] 
		\pgfplotstreampoint{\pgfpoint{3.92806cm}{-.0376cm}}
		\pgfplotstreamend
		\pgfusepath{stroke}
	 \end{tikzpicture}
	 \caption{\label{tricuts} Arcs of intersection}
	\end{center}
\end{figure}
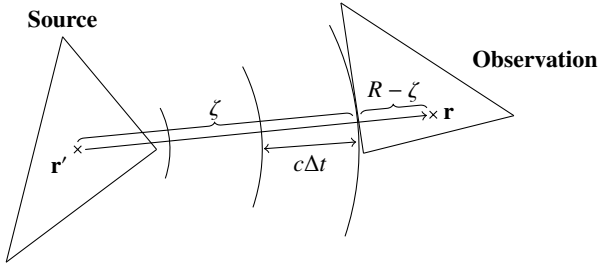

The approach espoused in this paper is to approximate the convolution of the space time basis function with the retarded potential. Note, that the TDEFIE in \eqref{eq:IntEq} requires both the convolution with the temporal derivative and the temporal integral of this space-time basis, and expressions for effecting these will be provided as we proceed. For simplicity, consider a point source and an observation triangle as shown in Fig. \ref{tricuts}. Consider a field due to the point source given by
\begin{equation}
\psi ({\bf r},t) = \frac{\delta\left(t - R/c\right )}{4 \pi R }\star_t T_i(t)~,
\end{equation}
where $\star_t$ denotes a convolution with respect to time. Given the piecewise nature of $T_i(t)$, it is evident that the lines of discontinuity correspond to intersections of time spheres with the observation triangle. This scheme has been implemented in acoustics \cite{Alte2003,Ha-Duong2003,Ha-Duong2003a} and in elecromagnetics \cite{Shanker2009} to great success. 

The approach proposed herein takes a slightly different path. Define the radii of the smallest and largest time spheres that enclose the triangle as $\left(\alpha c\Delta_t+\zeta\right)$ and $\left(\beta c\Delta_t+\zeta\right)$, respectively, where $\zeta$ is the largest multiple of $c\Delta t$ between the source point and observer triangle. 
Within this region, the convolution with the retarded potential can be expressed
as a separable expansion in space and time.  
Using this expansion, the field due to a point source located at $\br'$ with temporal dependence $T_i(t)$ may be approximated as 
\begin{equation} 
\label{eq:SepExp}
\begin{split}
\psi({\bf r},t) & =   \frac{1}{4 \pi R} \delta \left( t - \frac{\zeta}{c}\right) \star_t \delta \left ( t - \frac{R-\zeta}{c} \right ) \star_t T_i(t)~,\\ 
& = \frac{1}{4 \pi R} \delta \left (t - \frac{\zeta}{c}\right ) \star_t \sum_{l=0}^\infty a_l P_l \left(\hat{t}(R)\right) \tilde{T}_i^l(t)~,\\
& \approx \frac{1}{4 \pi R} \delta \left (t - \frac{\zeta}{c}\right ) \star_t \sum_{l=0}^{N_h} a_l P_l \left(\hat{t}(R)\right) \tilde{T}_i^l(t)~,\\
\text{where } &\hat{t}(R) = k_1 (R-\zeta)/c + k_2 ,~a_l = k_1\frac{2 l + 1}{2},\\&\tilde{T}_i^l(t)=P_l(k_1 t + k_2 ) {\cal P}_{\alpha,\beta}(t/\Delta_t) \star_t T_i(t)~.
\end{split}
\end{equation}
The constants $k_1$ and $k_2$ are chosen such that they map the domain $[\alpha, \beta] \rightarrow [-\Delta_t,\Delta_t]$, $P_l (\cdot)$ is a Legendre polynomial of order $l$ and $N_h$ is the number of harmonics that are retained in the expansion. 
The key consequence of this expression is the separation of space and time within the domain $R\in[\alpha c\Delta_t+\zeta,\beta c\Delta_t+\zeta]$.  As will become evident, this leads to spatial integrands which are smooth over the entire triangle.  It is evident that this can be generalized to the case of a source/observation pair. For instance, 
\begin{equation} 
	\begin{split}
		{\bf A}({\bf r},t)\circ \left\{ {\bf S}_m T_i \right \}  &= \frac{\mu_0}{4 \pi} \frac{\delta \left ( t - \frac{\left|{\bf r}\right|}{c} \right)}{\left|{\bf r}\right|} \star_{st} {\bf S}_m ({\bf r}) T_i(t)~,\\
		= \frac{\mu_0}{4 \pi} \int_{\Omega_m} d{\bf r}' &\frac{\delta \left ( t - \frac{R}{c} \right)}{R} \star_{t} {\bf S}_m ({\bf r}') T_i(t)~, \\
		= \frac{\mu_0}{4 \pi} \int_{\Omega_m} d{\bf r}' & \delta \left (t - \frac{\zeta}{c} \right)\\
		 \star_{t} ~& \frac{\delta \left ( t - \frac{R-\zeta}{c} \right)}{R}\star_t {\bf S}_m ({\bf r}') T_i(t)~. \\
	\end{split}
\end{equation}
Using \eqref{eq:SepExp} in the last term, it follows that 
\bml
\begin{equation}
\label{eq:ApotSep}
\left < {\bf S}_n ({\bf r}),{\bf A}({\bf r},t)\circ \left\{ {\bf S}_m T_i \right \}\right >  \approx \delta \left (t - \frac{\zeta}{c} \right) \star_t\frac{\mu_0}{4 \pi} \sum_{l=0}^{N_h} a_l \xi_l  \tilde{T}_i^l(t)~,
\end{equation}
where
\begin{equation}
	\displaystyle \xi_l  = \int_{\Omega_n} d{\bf r} {\bf S}_n ({\bf r})  \int_{\Omega_m} d{\bf r}' \frac{{\bf S}_m ({\bf r}')P_l \left(\hat{t}(R)\right)}{R}~. 
\end{equation}
\eml
Here $\zeta$, $\alpha$ and $\beta$ are chosen such that $\alpha \Delta_t < \left(R-\zeta\right)/c < \beta \Delta_t$ for all ${\bf r}' \in \Omega_m$ and ${\bf r} \in \Omega_n$. 
From the above expression it is evident that the spatial and temporal convolutions are completely independent of each other. It is also observed that the spatial convolution is smooth over the observation domain with a removable singularity. Given that one {\em a priori} knows $N_h$, integration rules can be designed to evaluate $\xi_l$ to high accuracy. The piecewise continuity or discontinuity of the temporal basis functions is present only in temporal convolutions and is handled relatively easily. 

\subsection{Implementation Details\label{Subsection:implementation}}

The above exposition details evaluation of the tested vector potential. Extensions to evaluate the specific components of \eqref{eq:MOT} are prescribed next. The matrix elements in \eqref{eq:MOT} comprise the temporal derivative of the vector potential and the scalar potential. Specifically, using \eqref{eq:ApotSep}, it follows that 
\begin{equation}
	\begin{split}
		&\left.\left < {\bf S}_n ({\bf r}) , \partial_t{\bf A}({\bf r},t) \circ \left \{ {\bf S}_mT_i \right \} \right>\right |_{ t= j \Delta_t}  =\\&~~ ~ ~ ~ ~ ~ ~ ~  ~ ~ ~ \delta \left (t - \frac{\zeta}{c} \right) \star_t\left.\frac{\mu_0}{4 \pi} \sum_{l=0}^{N_h} a_l \xi_l \partial_t\tilde{T}_i^l\right |_{t = j\Delta_t}
	\end{split}
\end{equation}
Note, the temporal basis function is compact. This convolution (at integer multiples of $\Delta_t$) can be evaluated analytically. The evaluation of the contribution due to the scalar potential is a little more involved. To this end, contribution of the scalar potential can be written as 
\bml
\begin{equation}
	\label{eq:phiexp}
	\begin{split}
		&\left.\left < {\bf S}_n ({\bf r}) , \nabla\Phi({\bf r},t) \circ \left \{ {\bf S}_mT_i \right \} \right > \right |_{ t= j \Delta_t}=\delta \left (t - \frac{\zeta}{c} \right) \star_t \\ & ~ ~ ~ ~ ~ ~ ~ \frac{1}{4 \pi\eps_0}  \sum_{l=0}^{N_h} a_l \tilde{\xi}_l P_l(k_1 t + k_2 ) {\cal P}_{\alpha,\beta}(t/\Delta_t) \star_t\\ &  ~ ~ ~ ~ ~ ~ ~ ~ ~ ~ ~ ~ ~ ~~ ~ ~ ~ ~ ~ ~\left. \int_{-\infty}^t dt' T_i (t')\right |_{ t= j \Delta_t}~,
	\end{split}
\end{equation}
where
\begin{equation}
	\label{eq:phiexp2}
	\begin{split}
\displaystyle \tilde{\xi}_l & = -\int_{\Omega_n} d{\bf r} \nabla\cdot{\bf S}_n ({\bf r})  \\ &~ ~ ~ ~ ~ ~ ~\int_{\Omega_m} d{\bf r}' \frac{\left(\nabla'\cdot{\bf S}_m ({\bf r}')\right)P_l \left(\hat{t}(R)\right)}{R}~.
\end{split}
\end{equation}
\eml
Here the vector derivative in ${\bf r}$ has been moved onto the testing function.  The evaluation of the convolution of the Legendre polynomial with the integral of the basis function may be written as 
\begin{equation}
\label{eq:convInt}
\begin{split}
P_l(k_1 t + k_2 ) {\cal P}_{\alpha,\beta}(t/\Delta_t) \star_t \int_{-\infty}^t dt' T_i (t') =&\\ \int_{-\infty}^\infty d\tau P_l (k_1 \tau + k_2) {\cal P}_{\alpha,\beta}(\tau/\Delta_t) \int_{-\infty}^{t - \tau} &dt' T_i (t')
\end{split}
\end{equation}
It can be shown that 
\begin{equation}
 \int_{-\infty}^{t-\tau}dt'T_i(t') = \begin{cases} {\displaystyle  \int_{-\infty}^{\infty}dt'T_i(t') } \mbox{ } t\geq \gamma \Delta_t \\
		{\displaystyle  \int_{-\infty}^{t-\tau}dt'T_i(t') } \mbox{ } t < \gamma \Delta_t
	\end{cases}~,
	\label{eq:convsplit}
\end{equation}
where $\gamma=\beta-\alpha+i+p$.  As a result, the integrals in \eqref{eq:convInt} decouple into a product of two integrals for $t \geq  \gamma\Delta_t$, and can be evaluated using the fact that $\int_{-\infty}^{\infty}d\tau P_l(k_1\tau+k_2) {\cal P}_{\alpha, \beta}(\tau/\Delta_t) =\delta_{0,l}(\beta-\alpha)\Delta t$, where $\delta_{i,j}$ is the Kronecker delta function. Using this, the values for sufficiently large $t$ can be moved to the right hand side of \eqref{eq:MOT}, analogous to the procedure in \eqref{eq:phisplit}.  The relation
\eqref{eq:oldchargecurrel} becomes
\begin{equation}
	\label{eq:chargecurrel}
  {\mathcal C}_j = ~{\mathcal C}_{j-1} + {\mathcal I}_{j-1} \int_{(j-2)\Delta t}^{(j+p-1)\Delta t} dt' T_{j-1}(t')
  \end{equation}
Finally, as an aside, we note a couple of other implementation features: 
\begin{enumerate} 
\item The temporal convolutions, while associated with the source and testing function, are independent of space. Furthermore, these are always tested using point testing at integer time steps. This implies that they need to be evaluated only once for each value of $\beta - \alpha$, and can be precomputed for a given geometry.
\item The highest order of the polynomial is known for a pair of basis functions and is given by $N_h$. As a result, rules for evaluating these integrals can be obtained from standard libraries or constructed. Furthermore, for self-triangles, the sinh$^{-1}$ rule outlined in \cite{Khayat2005} is used. Extension of this method to the integrand of each harmonic is trivial and is not elaborated here. 
\end{enumerate}

\section{Eigen Analysis of the MOT system \label{Section:eigen}}
In order to analyze the stability of the scheme proposed here, we perform an Eigen spectrum analysis similar to that in \cite{Walker2002}, but for the form of equations given here. Consider a simplification of the  marching scheme in equation \eqref{eq:MOT} and modified with the separable expansion, given by 
\bml
\begin{equation}
  {\mathcal Z}_0 {\mathcal I}_j = {\mathcal F}_j - \sum_{i=1}^{j-1} {\mathcal Z}_i {\mathcal I}_{j-i} -\sum_{i=1}^{j-1} \tilde{{\mathcal Z}}_i {\mathcal C}_{j-i}~,
  \label{eq:fullmot}
\end{equation} where ${\mathcal C}_j$ are the coefficients of the charge at time step $j$ and $\tilde{{\mathcal Z}}_i$ has been modified from that in \eqref{eq:phisplit} using the relation in \eqref{eq:convsplit}. From the equation of continuity we have that the charge, $\rho({\bf r}, t_i)$, is given by
\begin{equation}
	\begin{split}
  \rho({\bf r}, t_i) &= \int_{-\infty}^{t_i} dt' \nabla \cdot {\bf J}({\bf r},t')~,\\
  & =  \int_{-\infty}^{{t_i}-\Delta t} dt' \nabla \cdot {\bf J}({\bf r},t') + \int_{{t_i}-\Delta t}^{{t_i}} dt'\nabla \cdot {\bf J}({\bf r},t')~,\\
  & =  \rho({\bf r},{t_i}-\Delta t) +  \int_{{t_i}-\Delta t}^{{t_i}} dt'\nabla \cdot {\bf J}({\bf r},t')~,
	\end{split}
	\label{eq:cont}
\end{equation} 
\eml
which using the basis function expansions can be expressed as a relationship between the temporal coefficients of the charges and currents as defined in \eqref{eq:chargecurrel}. 
To start our analysis, we first note that the summations on the right hand side of equation \eqref{eq:fullmot} can be restricted to run from $i=1$ to $i=j-P$, where $P = \max(1,N_{max})$, where $N_{max}$, in turn, is the temporal extent of the scatterer. In other words, $(N_{max}-1)c\Delta t \leq diam(\Omega) \leq (N_{max})c\Delta t$, where $diam(\Omega)$ is the spatial extent of the scatterer along the incident field direction. 
Then equations \eqref{eq:fullmot} and \eqref{eq:chargecurrel} can  be written in matrix form as 
\begin{eqnarray} \left[ \begin{array}{c|c}
  \underline{A_{11}} & \underline{0} \\
  \underline{A_{21}} & \underline{\mathbb I} \\
  \end{array}\right] \underline{{\mathcal I}_j} = \underline{{\mathcal F}_j} 
  -  \left[ \begin{array}{c|c}
  \underline{B_{11}} & \underline{B_{12}} \\
  \underline{B_{21}} & \underline{B_{22}} \\
  \end{array}\right]\underline{{\mathcal I}_{j-1}}~, 
\label{eq:marchmat_z}
  \end{eqnarray}
where $\underline{A_{11}},~\underline{A_{21}},~\underline{B_{11}}
,~\underline{B_{21}},~\underline{B_{12}},~\underline{B_{22}}
,~\underline{{\mathcal I}_j},~\text{and}~\underline{{\mathcal F}_j}$
are defined in the appendix.
We denote the two matrices on either side of equation (\ref{eq:marchmat_z}) as $\underline{A}$ and $\underline{B}$ and assume that the temporal vector $\underline{{\mathcal I}_j}$ contains both current and charge coefficients. Assuming that the matrix $\underline{A}$ is invertible, and denoting $\underline{C} \doteq \underline{A}^{-1}\underline{B}$, the current vector at any given time step can be written as 
 \begin{eqnarray}
   \underline{{\mathcal I}_1} &=& \underline{A}^{-1} \underline{{\mathcal F}_1} - \underline{C} \underline{{\mathcal I}_{0}}~,\nonumber \\ 
   \underline{{\mathcal I}_2} &=& \underline{A}^{-1} \underline{{\mathcal F}_2} - \underline{C} \left\{\underline{A}^{-1} \underline{{\mathcal F}_1} - \underline{C}\underline{{\mathcal I}_{0}}\right\}~,\nonumber \\ 
   \underline{{\mathcal I}_3} &=& \underline{A}^{-1} \underline{{\mathcal F}_3} - \underline{C} \underline{A}^{-1} \underline{{\mathcal F}_2} + \underline{C}^2 \underline{A}^{-1} \underline{{\mathcal F}_1} - \underline{C}^3\underline{{\mathcal I}_{0}}~, \\ 
   &\ldots&\nonumber\\
   &\ldots& \nonumber\\
   \underline{{\mathcal I}_j} &=& \underline{A}^{-1} \underline{{\mathcal F}_j} + \sum_{k=0}^{j-1} (-1)^{k}\underline{C}^{k} \underline{A}^{-1} \underline{{\mathcal F}_k} +(-1)^j \underline{C}^j\underline{{\mathcal I}_{0}}~.\nonumber 
 \end{eqnarray}
 Now, let the eigenvalue decomposition of ${\mathcal C}$ be given by \cite{Golub1996} 
 \begin{subequations}
 \begin{equation}
   \underline{ {\mathcal C}} \doteq \sum_q \underline{\nu_q}^{\dagger} \sigma_q \underline{\nu_q}~, \label{eq:eigen}
 \end{equation}
 where $\dagger$ represents a conjugate transpose. Then, 
 \begin{equation}
   \underline{{\mathcal C}}^{k} \doteq \sum_q \underline{\nu_q}^{\dagger} \sigma_q^k \underline{\nu_q}
   \label{eq:eigenpower}
 \end{equation}
can be used to provide a natural bound on the matrix vector product as
 \begin{equation}
   \left\Vert  \underline{{\mathcal C}}^{k} \underline{{\mathcal F}} \right\Vert \leq \sigma_0^k \left\Vert \sum \underline{\nu_q}^{\dagger}\underline{\nu_q} \underline{ {\mathcal F}}\right\Vert  \label{eq:bnd1}
 \end{equation}
 \end{subequations}
 for any vector ${\mathcal F}$, where $\sigma_0$ is the largest eigenvalue of $\underline{{\mathcal C}}$. 
 Using the bound in \eqref{eq:bnd1}, ${\mathcal I}_j$ is bounded by 
 \begin{equation}
	 \begin{split}
   \left \Vert \underline{{\mathcal I}_j}\right \Vert \leq \left \Vert \underline{A}^{-1} \underline{{\mathcal F}_j} \right \Vert &+ 
   \sum_{k=0}^{j-1} \left\Vert \sigma_0^k  \sum_q \underline{\nu_q}^{\dagger}\underline{\nu_q} \underline{A}^{-1} \underline{{\mathcal F}_k} \right \Vert \\&+
   \left \Vert \sigma_0^j  \sum_q \underline{\nu_q}^{\dagger}\underline{\nu_q} \underline{{\mathcal I}_0} \right \Vert 
	 \end{split}
   \label{eq:fullbnd0}
 \end{equation}
 Assuming that after some time $j\Delta_t$ the input signal vanishes, i.e., ${\cal F}_j = 0 $, then the first summation on the right hand side of equation \eqref{eq:fullbnd0} can be restricted to $k\in{p_1,p_1+1,\dots,p_2}$ where the temporal extent of the input signal is from $p_1$ to $p_2$. Then \eqref{eq:fullbnd0} leads to \begin{equation}
    \left \Vert \underline{{\mathcal I}_j}\right \Vert \leq 
    \sigma_0^P P{\mathbb C}_1 +
   \sigma_0^j \left \Vert \sum_q \underline{\nu_q}^{\dagger}\underline{\nu_q} \underline{{\mathcal I}_0} \right \Vert ~,
   \label{eq:fullbnd1}
 \end{equation}
where $P = p_2-p_1 + 1$ and ${\mathbb C}_1 \doteq \left \Vert \sum_q \underline{\nu_q}^{\dagger} \underline{\nu_q} \max_k\left\{ \underline{A}^{-1} \underline{{\mathcal F}_{k}}\right\} \right \Vert$. At this point, it is apparent that depending on the value of $\sigma_0$, any numerical error in the initial current $\underline{{\mathcal I}_0}$ will grow without bound if $\sigma_0 > 1.0$, will decay to a value strictly bounded above by  $P{\mathbb C}_1$ is $\sigma_0 < 1.0$, and will have a constant DC value bounded above by $P {\mathbb C}_1$ if $\sigma_0 = 1$. 

 \section{Results\label{Section:results}}

 In this section, we present a number of results of scattering by perfect electrically conducting objects that are topologically very different from each other, some of which have, to the best of our knowledge, defied all stabilization attempts in the past with the exceptions of Lubich quadrature and exact integration or its variation \cite{Wang2008,Shanker2009,Shi2011}. Three pertinent features of the results presented are as follows: (i) the excitations are broadband, (ii) the solution time is very long (enough to permit multiple transits across the object), and (iii) some of the scatterers analyzed have features that make analysis difficult. The singular purpose of the objects chosen for analysis is to demonstrate late time stability for multiple transits on very challenging targets that are excited by a broad band pulse. In all cases, the incident field is a plane wave of the form 
\begin{equation}
	{\bf E}^{i}({\bf r},t)={\hat u}\cos(2\pi f_0t)e^{-(t-{\bf r}\cdot{\hat k} /c-t_p)^2/2\sigma^2 }~,
	\label{einc}
\end{equation}
where $\hat{u}$ denotes the polarization vector, $f_0$ the center frequency, 
${\hat k} =\sin(\theta^i)\cos(\phi^i)\hat{x}+\sin(\theta^i)\sin(\phi^i)\hat{y}+\cos(\theta^i)\hat{z}$ the direction of propagation, 
and $\theta^i$ and $\phi^i$ the polar and azimuthal angles of incidence, respectively.  
The values  $\sigma$ and $t_p$ are calculated as $\sigma=3/(2\pi B)$ and $t_p=6\sigma$, where $B$ denotes the
bandwidth of ${\bf E}^i$ in Hz.  The incident power is calculated to be approximately 160 dB below the peak
at $f_{max}=f_0+B$ and $f_{min}=f_0-B$.  For all results obtained in this paper, first order Lagrange polynomials are used as basis functions.  
In each of the examples, we present the current at a point on the geometry as a function of time. In addition, for some of the cases analyzed, we present an eigen spectra that shows that all values lie within the unit circle, with the exception of few that approach $1.0$. Finally, for some of the benchmark targets, we extract frequency domain radar scattering cross-section (RCS) data at three different frequencies, and this is then compared with similar data obtained either from an analytical code or a validated frequency domain integral equation solver. 

\subsection{Sphere}
\begin{figure}[!h]
     	\includegraphics[width=\columnwidth]{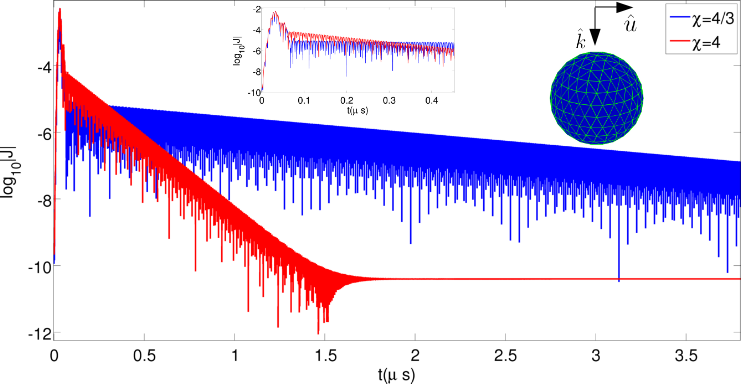}
     	\caption{Current on $1$ m radius sphere}
	\label{sphereCurr}
\end{figure}

Our first test case is a sphere of radius $1$ m discretized with $576$ unknowns.  It is excited with an 
incident wave propagating along $\hat{k}=\hat{z}$, polarized along $\hat{u}=\hat{x}$, with $f_{min}= 10$ kHz, and $f_{max}= 182$ MHz.  Shown in Fig. 
\ref{sphereCurr} is the current density at $(x,y,z)=(1,1.29,1.70)$ m for $47,000$  time steps, with $\chi=4/3$.  This example is also run for a much larger time step corresponding to $\chi = 4$. The inset shows the same current zoomed in the early time until .45 $\mu$s. As is evident from this figure, the currents are stable, expectedly so for the larger time step, as well as the smaller time step. The two currents agree well with each other. 
Fig. \ref{sphereCurr} shows that the surface current remains bounded over the duration of the simulation.  To show that
the result is indeed stable, we performed the analysis outlined in section \ref{Section:eigen}.  Fig. \ref{sphere_eig} shows that all of the eigenvalues lie within the unit circle and  as a result, stability can be expected. 
\begin{figure}[!h]
     \centering
     	\includegraphics[width=\linewidth]{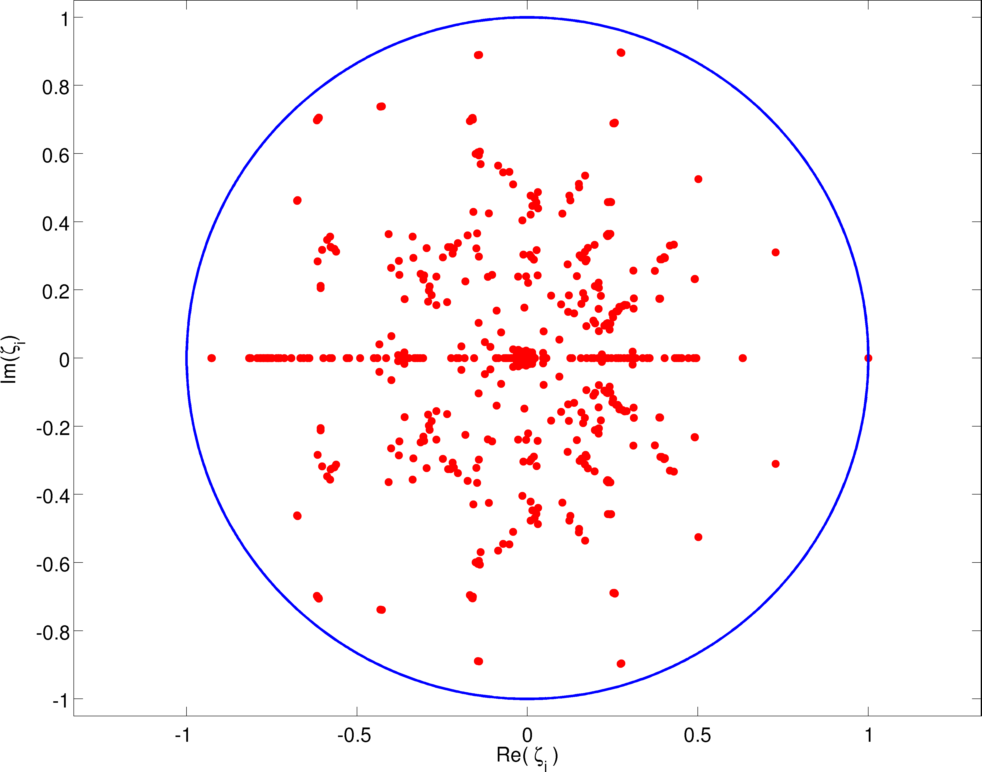}
     	\caption{Eigenvalues of sphere MOT matrix}
	\label{sphere_eig}
\end{figure}
Next, Fig. \ref{sphereRCS} compares the  RCS of the sphere, observed
in the $x-z$ plane for $\theta\in[-180\degree,180\degree]$, against that obtained using Mie series for three different frequencies. As is evident, agreement between the two sets of data is excellent. 
\begin{figure}[!h]
     \centering
     	\includegraphics[width=\linewidth]{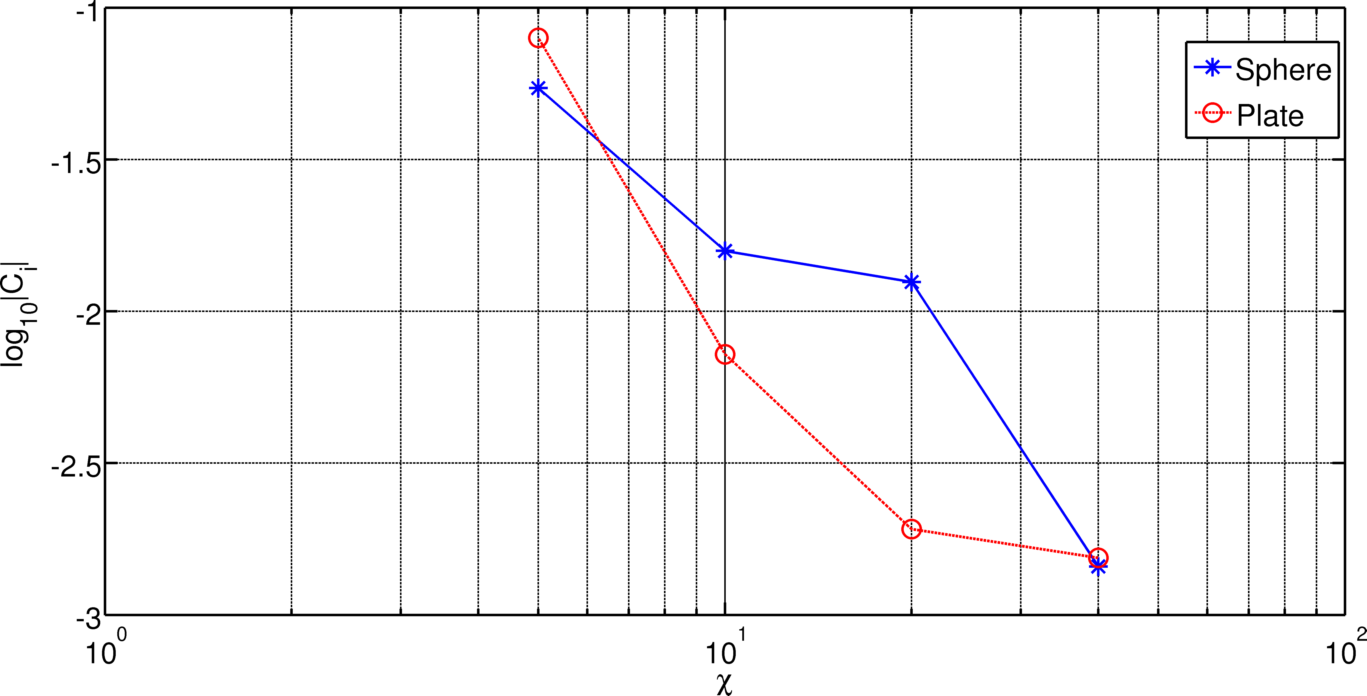}
     	\caption{Convergence of RCS of sphere and plate with respect to $\chi$}
	\label{sphere_eig}
\end{figure}
Fig. 5 shows the convergence with respect to oversampling factor $\chi$, of the RCS of the sphere at $f=f_0$ as well as a $1x1$ m plate discretized with
$133$ unknowns at $f=f_0=75$ MHz.  Convergence here is defined as $C_i=||\sigma(\chi_i)-\sigma(\chi_{i-1})||/||\sigma(\chi_i)||$ where, $||\cdot||$ is the $L_2$ norm and RCS $=10log_{10}|\sigma|$.  No more than 20 harmonics were used in any source/observer interactions in these 
simulations.
\begin{figure}[!h]
     \centering
     	\includegraphics[width=\linewidth]{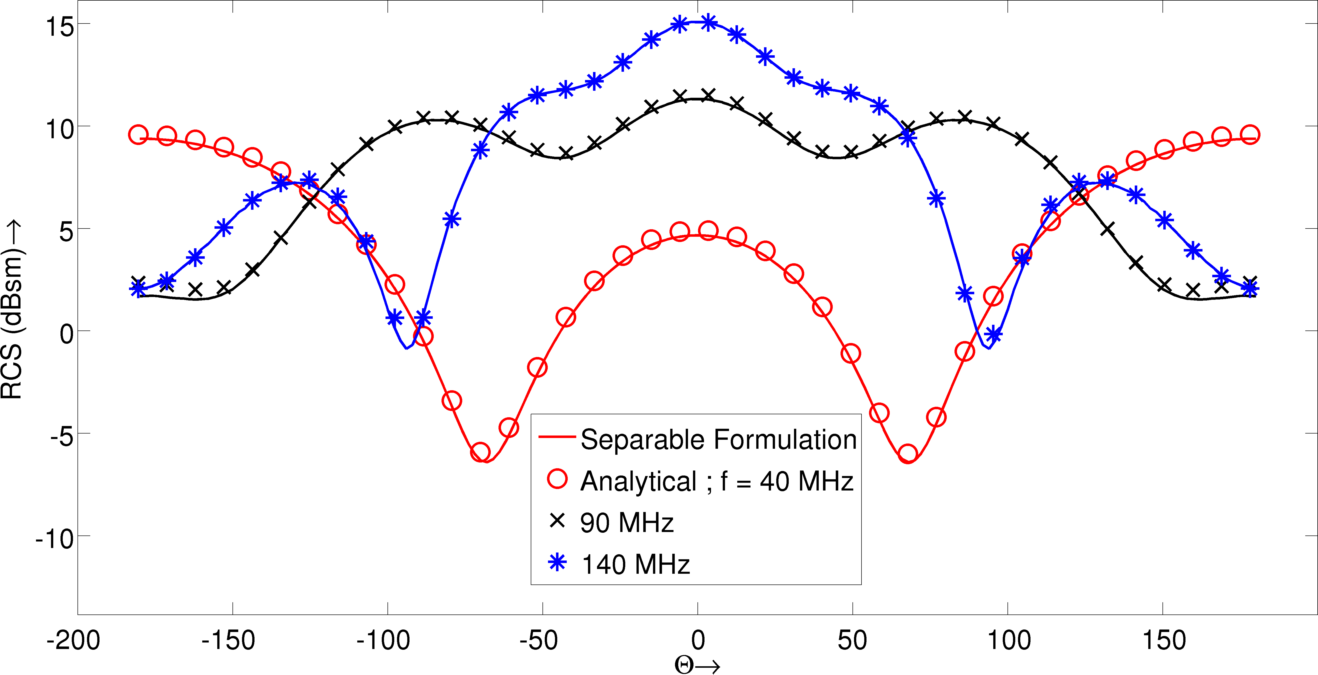}
     	\caption{RCS of $1$ m radius sphere}
	\label{sphereRCS}
\end{figure}

\subsection{Two Plates}

\begin{figure}[!h]
     \centering
     \includegraphics[width=\linewidth]{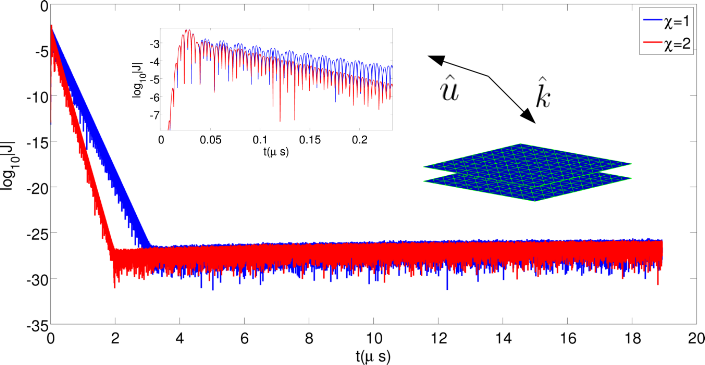}
     \caption{Current on two parallel plates}
	\label{twoPlateCurr}
\end{figure}

Next, we apply this scheme to analyze scattering from an object described using an open surface. Here, the object chosen was similar to that in \cite{Shi2011} and consists of two plates of size 1m $\times$ 1m parallel to the $x-y$ plane that are separated by $0.1$m. Each plate is discretized using 560 spatial unknowns. The incident field is polarized along $\hat{u}=\frac{1}{2}(\hat{x}+\hat{y}-\sqrt2\hat{z})$, propagates along $\hat{k}=\frac{1}{2}(\hat{x}+\hat{y}+\sqrt2\hat{z})$, and is described by parameters $f_{min}= 100$ kHz and $f_{max}= 264$ MHz. The current is observed at $(x,y,z)=(0.95,0.95,0.1)$ m  for $100,000$ time steps for a time step given by $\chi =1 $. This corresponds to approximately 3,383 transits across the object after the incident field has died down. Similar data is obtained for time step size corresponding to $\chi = 2$ for 50,000 time steps. As is evident from Fig. \ref{twoPlateCurr}, the currents are stable for the entire duration.

\subsection{More challenging targets}

In this section, we present results for three challenging targets, a thin box, an even thinner NASA almond, and a cone sphere. All these scatterers are illuminated by a broadband pulse and analyzed for multiple transits across their surfaces. 

\subsubsection{Thin Box}
\begin{figure}[!h]
     \centering
     \includegraphics[width=\linewidth]{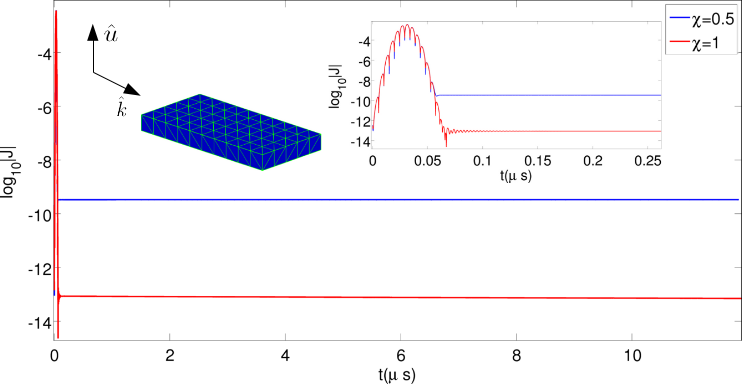}
     \caption{Current on thin box}
	\label{thinBoxCurr}
\end{figure}

\begin{figure}[!h]
     \centering
     \includegraphics[width=\linewidth]{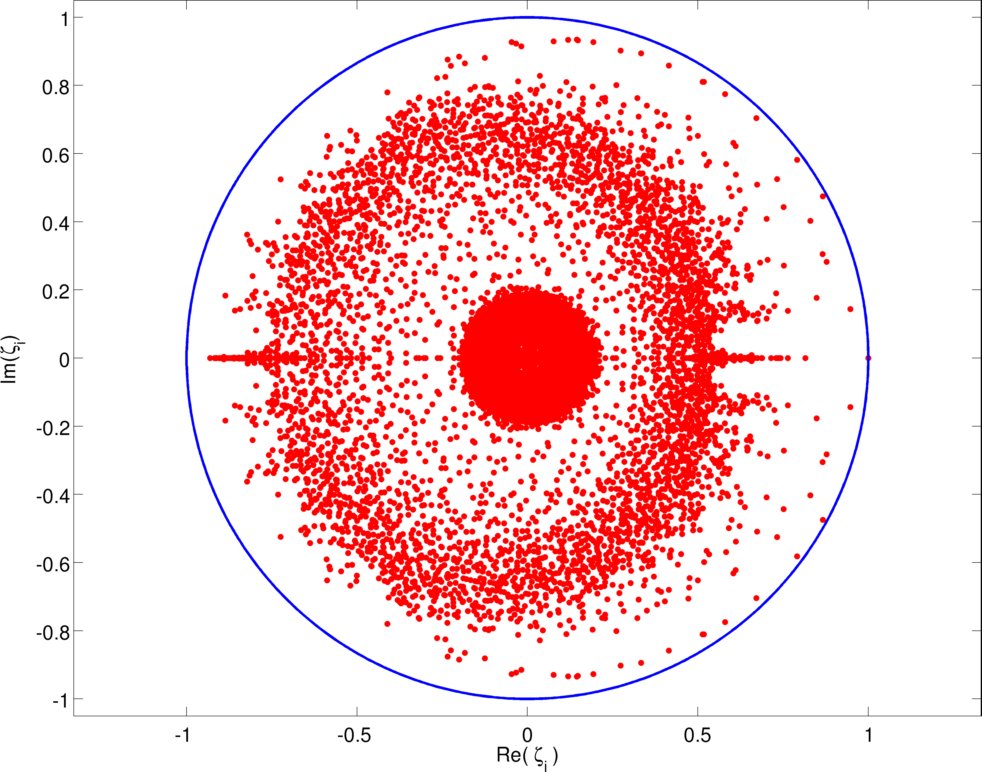}
     \caption{Eigenvalues of thin box MOT matrix}
	\label{thboxeig}
\end{figure}

Next, we analyze scattering from a thin box; again as mentioned earlier, this has been a challenge insofar as stability of the TDEFIE is concerned. To the best of our knowledge, stable results have been obtained only using exact integration and variations thereof \cite{Shanker2009,Shi2011}. The dimension of the box are 0.5m $\times$ 1m $\times$ 0.1m and is discretized with 390 spatial unknowns.  The box is excited by a field that is propagating along $\hat{k}=-\hat{y}$, polarized along $\hat{u}=\hat{z}$, and described by parameters $f_{max}= 211$ MHz and $f_{min}= 1$ MHz. The current is observed at $(x,y,z)=(0.5,0,0)$ m, for 50,000 time steps for a time step size given by $\chi = 0.5$ and for 25,000 time steps for a time step size given by $\chi = 1$. As is evident from Fig. \ref{thinBoxCurr}, the two currents agree well with each other and they are stable. In addition to a full MOT solution, we conducted an eigen-analysis of the MOT system of equations, and as is evident from Fig. \ref{thboxeig}, all the eigenvalues lie within the unit circle. This again attests to the stability of the MOT system for this scatterer. 

\subsubsection{NASA Almond}

\begin{figure}[!h].
     \centering
     \includegraphics[width=\linewidth]{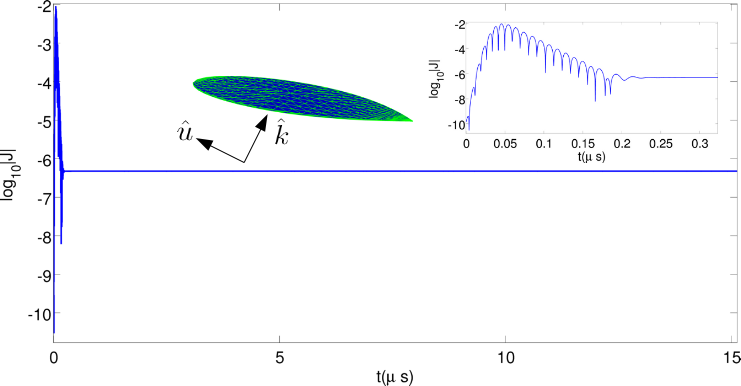}
     \caption{Current on NASA almond}
	\label{almondCurr}
\end{figure}

\begin{figure}[!h]
     \centering
     \includegraphics[width=\linewidth]{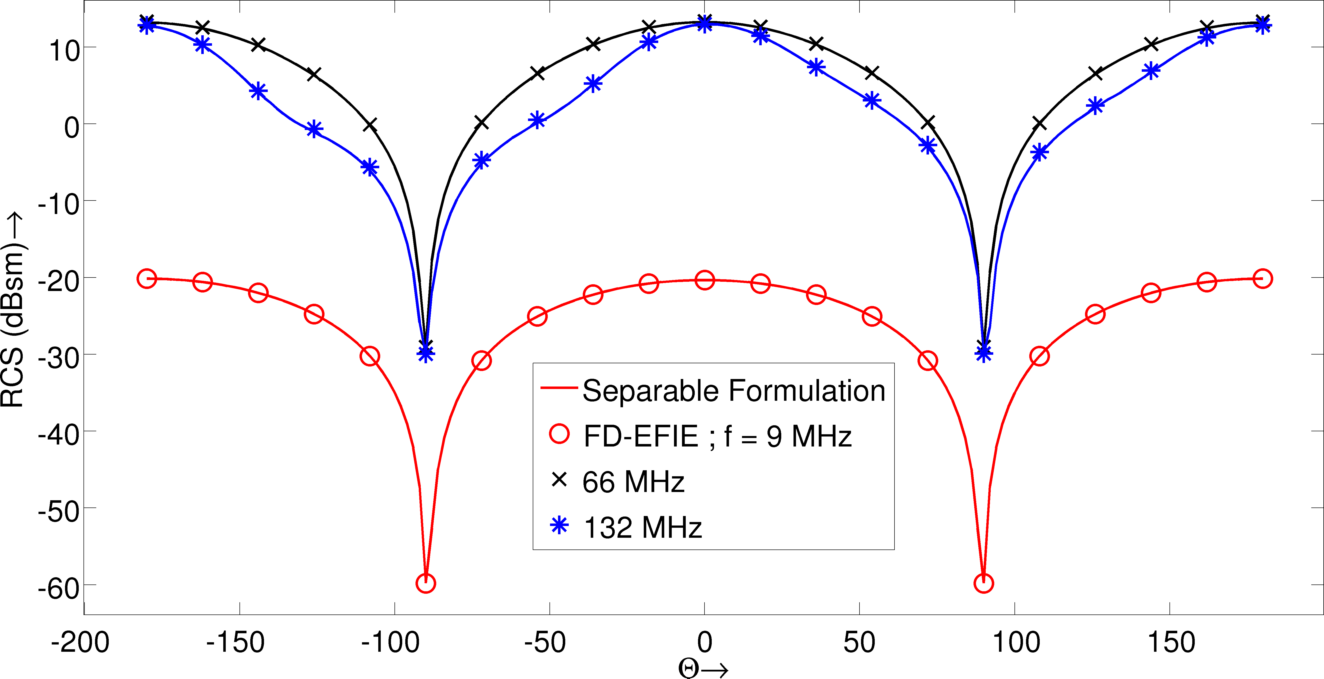}
     \caption{RCS of NASA almond}
	\label{almondRCS}
\end{figure}

Next, we analyze scattering from a thin almond; the almond fits in a box of dimension 2.17m $\times$ 1.11m $\times$ 0.06m (aspect ratio 38:20:1), and is discretized using 1140 unknowns. The target is illuminated by a field incident along $\hat{k}=\hat{z}$, polarized along $\hat{u}=-\hat{x}$, and characterized by $f_{max}=132$ MHz and $f_{min}=1$ kHz. The current at $(x,y,z)=(38.0,689,28.8)$ for 40,000 time steps with size determined by $\chi = 1$ is shown in Fig. \ref{almondCurr}. As is evident from this figure, the current does not show late time instability. The inset in Fig. \ref{almondCurr} shows the features of the current until $t = 0.3 \mu s$. Next, we extract RCS data at three different frequencies, in the $x-z$ plane for $\theta\in[-180\degree,180\degree]$,  and compare these results against similar data obtained using a frequency domain code. Again, as is evident in Fig. \ref{almondRCS}, the agreement between the two sets of data is excellent. 
\subsubsection{Cone-Sphere}
\begin{figure}[!h]
     \centering
     \includegraphics[width=\linewidth]{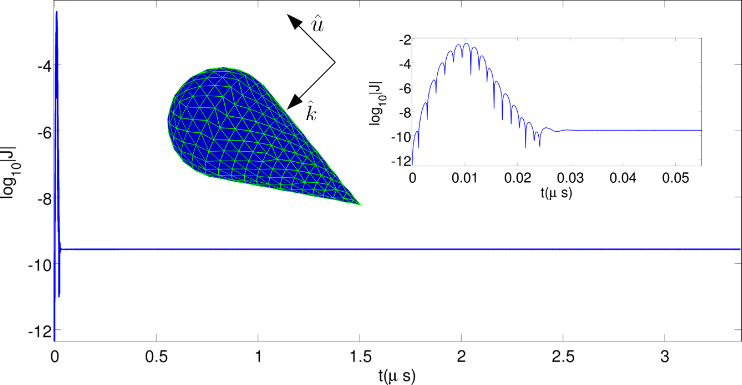}
     \caption{Current on cone-sphere}
	\label{cnSphereCurr}
\end{figure}

\begin{figure}[!h]
     \centering
     \includegraphics[width=\linewidth]{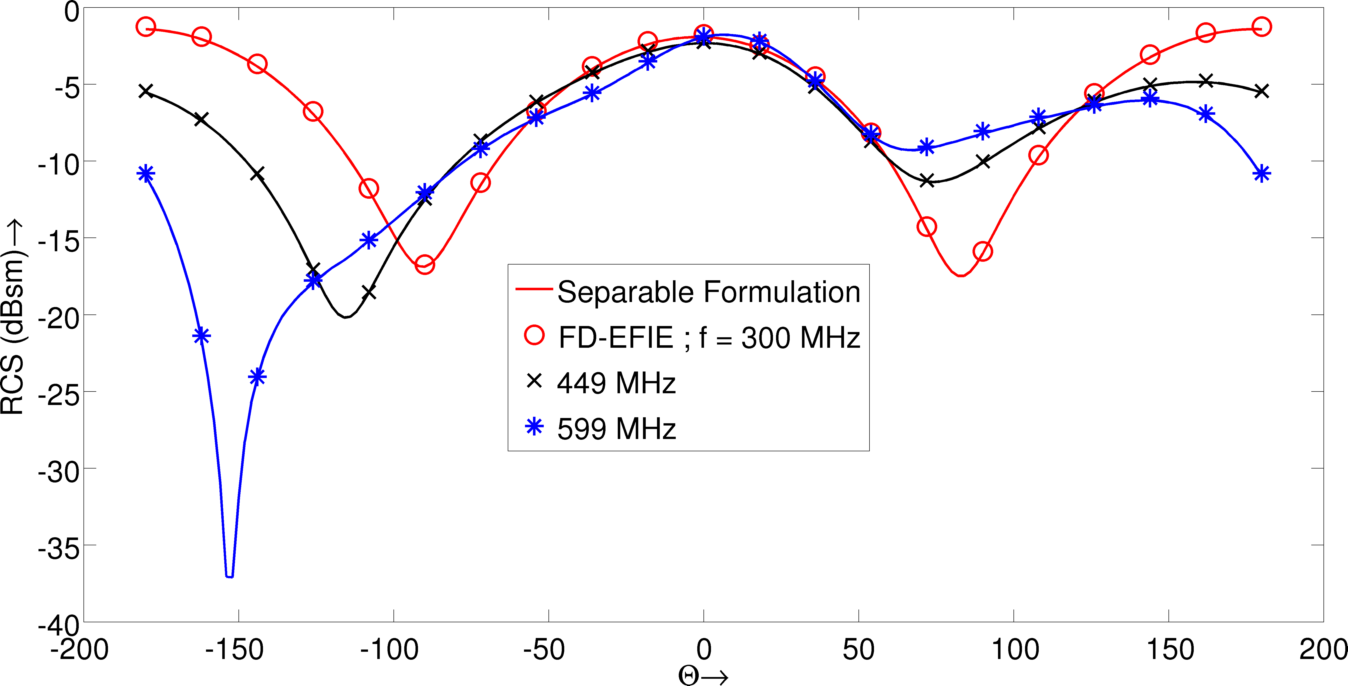}
     \caption{RCS of cone-sphere}
	\label{cnSphereRCS}
\end{figure}

In this last example, we analyze scattering from a cone-sphere. The radius of the base of the cone is $0.25$m while its height is $1$m, and the scatterer is represented using 1008 spatial degrees of freedom. The incident field is propagating along $\hat{k}=\hat{z}$, is polarized along $\hat{u}=\hat{x}$, and is characterized by $f_{max}=599$ MHz and $f_{min}=1$ MHz. The current observed at $(x,y,z)=(0.329,0.335,0.125)$ m for 40,000 time steps with the time step size corresponding to $\chi = 1$ is depicted in Fig. \ref{cnSphereCurr}. As is evident from this figure, the currents exhibit late time stability for multiple transits across the geometry. An inset in Fig. \ref{cnSphereCurr} depicts features until $t = 55 ns$. As before, RCS data in the $x-z$ plane for $\theta\in[-180\degree,180\degree]$ is extracted at three different frequencies and compared against similar data obtained using a frequency domain code. As is evident from Fig. \ref{cnSphereRCS}, the agreement is excellent at all three frequencies. 

\section{Conclusion\label{Section:summary}}
This paper presents a novel framework for constructing TDIEs; the crux of the approach presented in this paper lies in developing a separable spatio-temporal expansion for representing the convolution between the retarded potential and the source. As a result of this expansion, the discontinuities in the temporal basis set do not appear in the spatial integrands. This method, in concert with the correct variational form, has been applied to the analysis of scattering  from number of challenging targets for multiple transits of the incident pulse across the object. The goal of this set of experiment was two-fold: (i) study stability behavior in early time, and (ii) analyze behavior when the incident field has completely died down. In all cases, the currents reach a DC floor, and remain there for the duration of the analysis; the DC floor is expected as it lies in the null space of the TDEFIE operator.   These results demonstrate the viability of using this approach for TDIE analysis, and opens door to more challenging analysis. Extension of this approach to higher order geometries as well as integration with PWTD accelerators is underway and will be presented elsewhere. 

\section{Appendix}
\begin{subequations}
\begin{eqnarray}
  \underline{A_{11}}=&\left[ \begin{array}{ccccc}
-\left[{\mathcal Z}_0\right] & 0 &   & &   \ldots \\
0 &\left[{\mathbb I}\right] & 0 &    & \ldots \\
0 & 0 &\left[{\mathbb I}\right] & 0   & \ldots \\
. & & &  & \\
\ldots & & &  & \left[{\mathbb I}\right]
  \end{array}\right]~,\\
  \underline{A_{21}}=&\left[ \begin{array}{ccccc}
-\left[{{\mathbb T}_j}\right]  & 0 &   & &   \ldots \\
0 & 0 &  &    & \ldots \\
 \ldots &  & &   & \ldots \\
\ldots & & &  & 0
  \end{array}\right]~,\\
  \underline{B_{11}}=&\left[ \begin{array}{cccc}
\left[{\mathcal Z}_1\right] & \left[{\mathcal Z}_2\right] & \ldots &  \left[{\mathcal Z}_P\right] \\
  \left[{\mathbb I}\right] &0 &    & \ldots \\
   0 &\left[{\mathbb I}\right]  &  & \ldots \\
 . & &  & \\ 
 . & &  & \\
 \ldots & &\left[{\mathbb I}\right]&0 
  \end{array}\right]~,\\
  \underline{B_{12}}=&\left[ \begin{array}{ccccc}
 \left[\tilde{ \mathcal Z}_1\right] & \left[\tilde{\mathcal Z}_2\right] & \ldots &   & \left[\tilde{\mathcal Z}_P\right]\\
   \left[{\mathbb I}\right] &0 &  &    & \ldots\\
   \left[{\mathbb I}\right] & 0 &  &    & \ldots\\
   . & &  & & \\
   . & &  & & \\
   \ldots &  & &\left[{\mathbb I}\right]&0
   \end{array}\right]~,\\
  \underline{B_{21}}=&\left[ \begin{array}{cccc}
\left[{{\mathbb T}_{j-1}}\right] & 0 &\ldots  & \ldots \\
 0 &0  &  & \ldots\\
    . & & & \\
   . & & & \\
    \ldots & &0 &0  \\
   \end{array}\right]~,\\
  \underline{B_{22}}=&\left[ \begin{array}{ccccc}
 \left[{\mathbb I}\right] & 0 &  &    & \ldots\\
  \left[{\mathbb I}\right] & 0 &  &    & \ldots\\
  . & &  & & \\
  . & &  & &\\
  \ldots &  & &\left[{\mathbb I}\right]&0
  \end{array}\right]~,
\end{eqnarray}
\begin{equation}
  {\mathbb T}_j \doteq  \int_{(j-p-1)\Delta t}^{j\Delta t} dt' T_{j}(t')~,\\
\end{equation}
and \\
\end{subequations}
where $\left[{\mathbb I}\right]$ is the identity matrix.  
\section{Acknowledgments}

The authors would like to acknowledge computing support from the HPC Center at Michigan State University, financial support from NSF via DMS 0811197 and CCF 1018516. 

\bibliographystyle{ieeetr}
\bibliography{bigbib}

\end{document}